\documentclass[aps,pre,twocolumn,showpacs,superscriptaddress,groupedaddress]{revtex4-1}  

\usepackage{hyperref}
\usepackage{graphicx}  
\usepackage{dcolumn}   
\usepackage{bm}        
\usepackage{amssymb}   
\usepackage{color}     

\begin{document}
\title{Two-stage melting induced by dislocations and grain boundaries in monolayers of hard spheres}
\author{Weikai Qi}
\affiliation{Soft Condensed Matter, Debye Institute for Nanomaterials Science, Utrecht University, Princetonplein 5, 3584 CC Utrecht, The Netherlands}

\author{Anjan P. Gantapara}
\affiliation{Soft Condensed Matter, Debye Institute for Nanomaterials Science, Utrecht University, Princetonplein 5, 3584 CC Utrecht, The Netherlands}

\author{Marjolein Dijkstra}
\email{M.Dijkstra1@uu.nl}
\affiliation{Soft Condensed Matter, Debye Institute for Nanomaterials Science, Utrecht University, Princetonplein 5, 3584 CC Utrecht, The Netherlands}
\date{\today}

\begin{abstract}
Melting in two-dimensional systems has remained controversial as theory, simulations, and experiments show contrasting results. One issue that obscures this discussion is whether or not theoretical predictions on strictly 2D systems describe those of quasi-2D experimental systems, where out-of-plane fluctuations may alter the melting mechanism.
Using  event-driven Molecular Dynamics simulations,  we find that the peculiar two-stage melting scenario of a continuous solid-hexatic  and a first-order hexatic-liquid transition as observed for a {\em truly 2D} system of hard disks [Bernard and Krauth, Phys. Rev. Lett. {\bf 107}, 155704 (2011)] persists for a quasi-2D system of hard spheres with out-of-plane particle motions {\em as high as} half the particle diameter. By calculating the renormalized Young's modulus, we show that the  solid-hexatic transition is of the Kosterlitz-Thouless type, and occurs  via dissociation of bound dislocation pairs. In addition, we find a first-order hexatic-liquid transition that seems  to be driven by a spontaneous proliferation of grain boundaries.
\end{abstract}


\maketitle

\section{Introduction}
Melting in two-dimensional (2D) systems has been debated heavily since Landau, Peierls, and Mermin showed that thermal long-wavelength fluctuations do not allow for long-range positional order in a 2D solid ~\cite{RevModPhys.60.161, RevModPhys.71.1737, 0953-8984-21-20-203101}. According to the Kosterlitz-Thouless-Halperin-Nelson-Young (KTHNY) theory, the melting mechanism of 2D crystals proceeds via two consecutive continuous transitions, which are induced by the formation of topological defects~\cite{KT73, Nelson79,Young79}. A topological defect in a two-dimensional crystal with triangular symmetry is defined as a particle that does not possess six  nearest neighbors: a disclination is an isolated defect with five or seven nearest neighbours, while a dislocation is an isolated pair of a 5- and 7-fold defect. According to the KTHNY theory, the 2D solid melts  via dissociation of  bound dislocation pairs ($5-7-5-7$ quartets) into an intermediate hexatic phase, which is characterized by short-ranged  positional order, but quasi-long-ranged  bond orientational order. Subsequently, the hexatic phase transforms into a liquid phase with short-ranged positional and orientational order via the unbinding of dislocations ($5-7$ pairs) into separate disclinations~\cite{KT73, Nelson79,Young79}. However, the KTHNY theory only predicts when the system becomes unstable with respect to unbinding of dislocations and disclinations, and does not rule out the possibility that these transitions might be preempted by a single first-order fluid-solid transition \cite{binder_2002} driven  by an alternative melting mechanism, e.g., grain-boundary induced melting ~\cite{chui_prl_1982,saito_prl_1982}. Many computer simulation and experimental studies have been performed to reveal the melting mechanism for 2D solids with contrasting support  for  a two-stage melting scenario via an hexatic phase as well as  a first-order melting transition~\cite{PhysRevLett.82.2721,karnchanaphanurach_pre_2000,PhysRevLett.104.205703,PhysRevE.77.041406,rice_CPL_2009,murray_prl_1987,kusner_prl_1994,mazars,marcus_prl_1996,grunberg_prl_2004,keim_pre_2007,deutchlander_prl_2013,lin_jcp_2007}.  These results seem to suggest that 2D melting is not universal, but depends on
specific properties of the system, \emph{e.g.}, interparticle potential, out-of-plane fluctuations, finite-size effects, etc.

For a 2D hard-disk system, important progress has been made recently as large-scale simulations confirmed the existence of an hexatic phase, but found in contrast to predictions of the KTHNY theory a first-order liquid-hexatic phase transition  and  a continuous hexatic-solid transition \cite{Jaster2004120,PhysRevLett.107.155704}. These results,  confirmed by three different simulation methods in Ref.~\cite{arXiv1211.1645},  settled a long-standing debate on the nature of 2D hard-disk melting, which was fueled by conflicting results mainly caused by finite-size effects and poor statistics due to insufficient computer power in previous studies~\cite{PhysRev.127.359,hoover_jcp_1968,lee_prb_1992,PhysRevB.46.11186,weber_prb_1995,PhysRevE.59.2659,0295-5075-27-8-007,PhysRevLett.75.3477,PhysRevE.55.6855,fernandez_pre_1997,Jaster_pre_1999,PhysRevE.73.065104}.

In order to study 2D melting in experiments \cite{rice_CPL_2009}, colloidal particles were confined between two glass plates, showing observations consistent with the KTHNY scenario  \cite{PhysRevE.77.041406,murray_prl_1987,kusner_prl_1994}, a first-order fluid-solid transition \cite{karnchanaphanurach_pre_2000}, and a first-order liquid-hexatic and first-order hexatic-solid transition \cite{marcus_prl_1996}. It is important to note that in these experiments the separation between the glass plates were on the order of 1.2 to 1.5 times the particle diameter, and hence the particles can move out of plane. An alternative way to study 2D melting in colloidal systems is to adsorb the particles at air-liquid or liquid-liquid
interfaces, which restricts significantly the out-of-plane motion of the particles. Support has been found in these
experimental set-ups for the two-stage KTHNY melting  \cite{PhysRevLett.82.2721,grunberg_prl_2004,keim_pre_2007,deutchlander_prl_2013}, but also for a first-order fluid-hexatic and continuous hexatic-solid transition  \cite{lin_jcp_2007}.
Clearly, there is no consensus in simulations and experiments on the nature of the 2D melting transition. These results may suggest that the melting mechanism depends sensitively on the interparticle interactions. However, even for particle systems interacting with short-range repulsive pair potentials conflicting results have been found experimentally \cite{PhysRevE.77.041406,murray_prl_1987,karnchanaphanurach_pre_2000}. It therefore remains essential to investigate
whether or not  out-of-plane fluctuations can alter the melting scenario, and can explain the
discrepancies in the experimental results.

In this paper, we focus  on particle systems that interact with  excluded-volume interactions. To be more precise, we investigate the effect of out-of-plane fluctuations  on the melting mechanism of hard spheres confined between two parallel hard plates using event-driven Molecular Dynamics (EDMD) simulations. We find that the peculiar melting mechanism of a  quasi-2D monolayer of hard spheres is very similar to that of a 2D hard-disk system~\cite{PhysRevLett.107.155704,arXiv1211.1645} even
when the out-of-plane fluctuations are as large as half a particle diameter, and thus experiments on tightly confined colloids should show  a continuous hexatic-solid transition and a first-order hexatic-liquid transition provided the interactions are hard-sphere-like. This result is highly surprising as a previous simulation study of Lennard-Jones particles shows that the hexatic phase disappears when the particles undergo tiny out-of-plane fluctuations \cite{gribova:054514}. Similarly, simulations on attractive colloidal soft spheres show that the KTHNY melting transition in 2D systems  can change to a first-order transition in quasi-2D systems with out-of-plane fluctuations of 1.2 $\sigma$~\cite{frydel_2003,rice_CPL_2009}. More importantly, we also provide an explanation for the observed melting behavior.  By calculating the  renormalized Young's modulus   for the solid phase, we show that the solid-hexatic transition is of the Kosterlitz-Thouless (KT) type, and is driven by   the formation of isolated dislocations. However, the melting of the hexatic phase proceeds via a first-order grain-boundary induced melting transition that intervenes the KTHNY scenario.


\begin{figure*}
\centering
\includegraphics[width=0.85\textwidth]{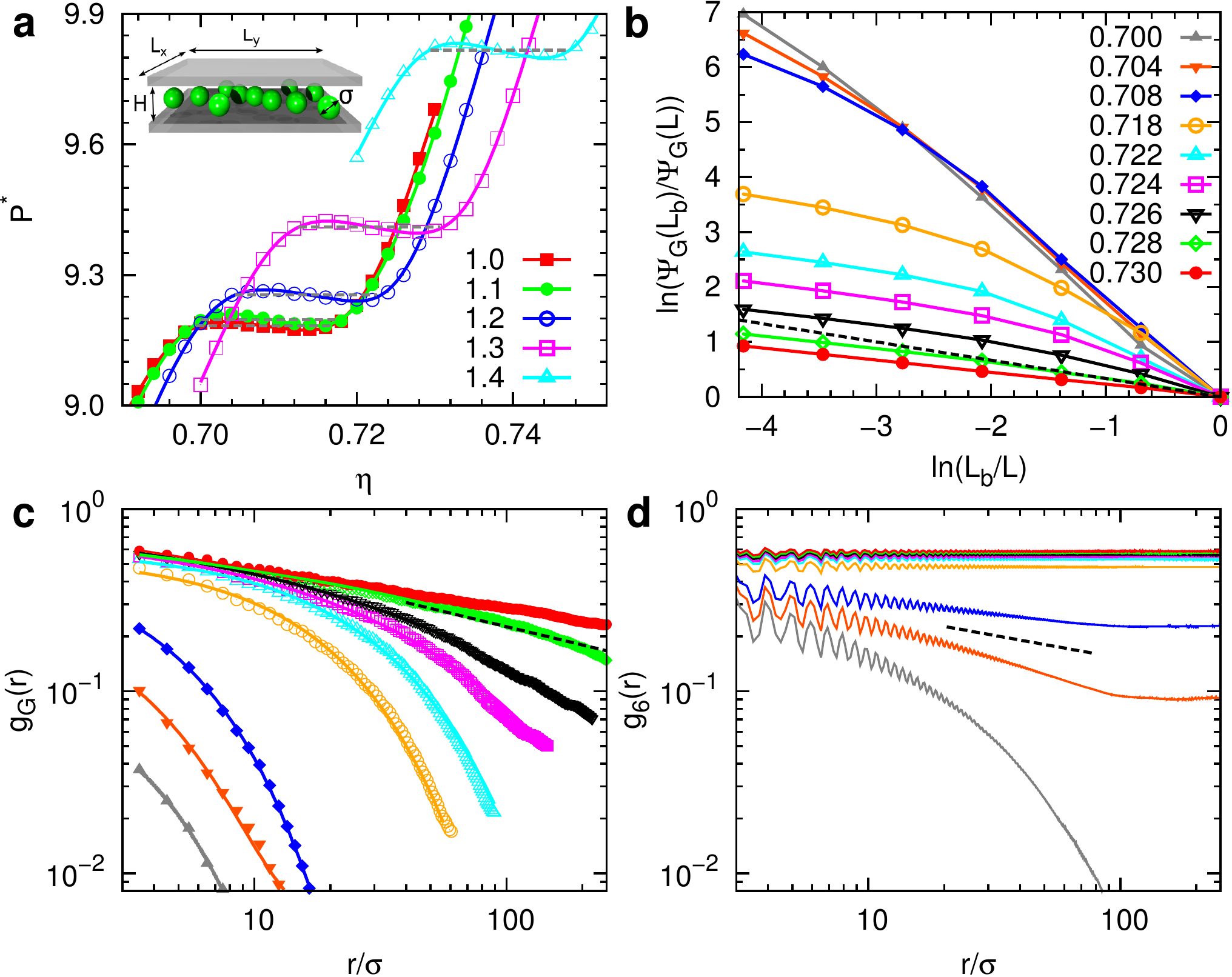}
\label{Fig:Scalinghd}
\caption{(a) The reduced 2D lateral pressure $P^*=\beta P \sigma^2$ as a function of the 2D packing fraction $\eta=\pi N \sigma^2/4A$ for $N=1024^2$ hard spheres with diameter $\sigma$ confined between two parallel hard walls with varying  plate separation $H/\sigma$ as labeled and area $A = L_x L_y$ as illustrated in the inset. The gray dashed lines indicate the coexistence regions as obtained from the Maxwell construction.  (b) Subblock  scaling analysis of the 2D  positional order parameter in reciprocal space $\Psi_G(L_b)$ versus $L_b$ for $H/\sigma=1.1$ for varying $\eta$ as labeled. (c)
Positional correlation function $g_{\mathbf{G}}(r)$  and (d) bond orientational order correlation function $g_6(r)$ as a function of $r$ for $H/\sigma=1.1$ and varying $\eta$ as labeled in (b). The slope of the black dashed line in (b) and (c) corresponds to $-1/3$, which indicates a hexatic-solid transition at $\eta_{HS} \simeq 0.728$ according to the KTHNY theory. The slope of the black dashed line in (d) is $-1/4$, which equals the maximum possible slope for a hexatic phase.  The solid lines in (c) are fits of $g_G(r)$.}
\end{figure*}

\section{Model and Methods}

We performed large-scale event-driven Molecular Dynamics (EDMD) simulations of  $N=1024^2=1,048,576$ hard spheres with diameter $\sigma$ confined between two parallel hard plates of area $A = L_x L_y$ with $L_x : L_y=2 : \sqrt{3}$ to accomodate a crystalline layer with triangular symmetry, as illustrated in the inset of Fig. 1(a). In an EDMD simulation, the system evolves via a time-ordered sequence of elastic collision events, which are described by Newton's equations of motion.  The spheres move at constant velocity between  collisions, and  the velocities of the respective particles are updated when a collision occurs. All collisions are elastic and preserve energy and momentum.  In order to speed up the equilibration we divided the simulation box into small cells in the XY plane, and we used a cell list~\cite{b3187662}. In addition, we employed an event calendar to  maintain a list of all future events~\cite{b3187662}. Three different events are listed in the calendar: (1) collisions between particles; (2) collisions between particles with the two  walls; and (3) particles that cross a cell boundary.

The phase behavior of this system was  determined as a function of plate separation $H$ in Refs.~\cite{PhysRevLett.76.4552,0953-8984-18-28-L02,franosch_2012}. The phase diagram as determined from free-energy calculations shows a first-order phase transition from a fluid phase to a crystal phase consisting of a single triangular layer  for plate separations $1 \leq H/\sigma \leq 1.53$~\cite{0953-8984-18-28-L02}. However, the presence of an intermediate hexatic phase was ignored in this study. We also note that  the system reduces to a 2D system of hard disks for $H/\sigma=1$.

\section{Results}
\subsection{Mayer-Wood loop in the equation of state}
We performed EDMD simulations in the $NVT$ ensemble for varying plate separations $1 \leq H/\sigma \leq 1.53$. We computed the reduced 2D lateral pressure $P^*$ from the collision rate via the virial theorem given by
\begin{equation}
 P^*= \beta P \sigma^2 = \frac{N \sigma^2}{A}\left[1-\frac{\beta m}{2t}\frac{1}{N}\sum_{i<j}^N{\mathbf{r}_{ij} \cdot \mathbf{v}_{ij}} \right],
\end{equation}
 where $m=1$ is the mass of the particles, $\beta=1/k_BT$ the inverse temperature, $k_B$ Boltzmann's constant, $t$ is the time interval, $\mathbf{r}_{ij}$ and $\mathbf{v}_{ij}$ are the 2D projections of the distance vector and the velocity vector, respectively, between particle $i$ and $j$.

In Fig. 1(a), we plot $P^*$ as a function of the 2D packing fraction $\eta=\pi N \sigma^2/4A$ for varying  plate separations $1 \leq H/\sigma \leq 1.53$. For all $H/\sigma$ considered, we observed  a Mayer-Wood loop in the equation of state (EOS) due to interfacial tension effects in finite systems~\cite{mayer_jcp_1965}. We determine the coexisting densities using a Maxwell construction as presented in the Supplementary Information. The presence of such a loop in the EOS provides  support for a first-order phase transition. We note however that such loops in the EOS can also appear due to the finite size of 2D systems \cite{PhysRevE.59.2659}. We therefore also verified that the interfacial free energy $f$ obtained from integrating the EOS scales as $f \propto N^{-1/2}$, which yields strong evidence for a first-order transition from an isotropic fluid (also referred to as a liquid) phase to a more ordered phase~\cite{PhysRevLett.107.155704}.

\begin{figure}
\centering
\includegraphics[width=0.5\textwidth]{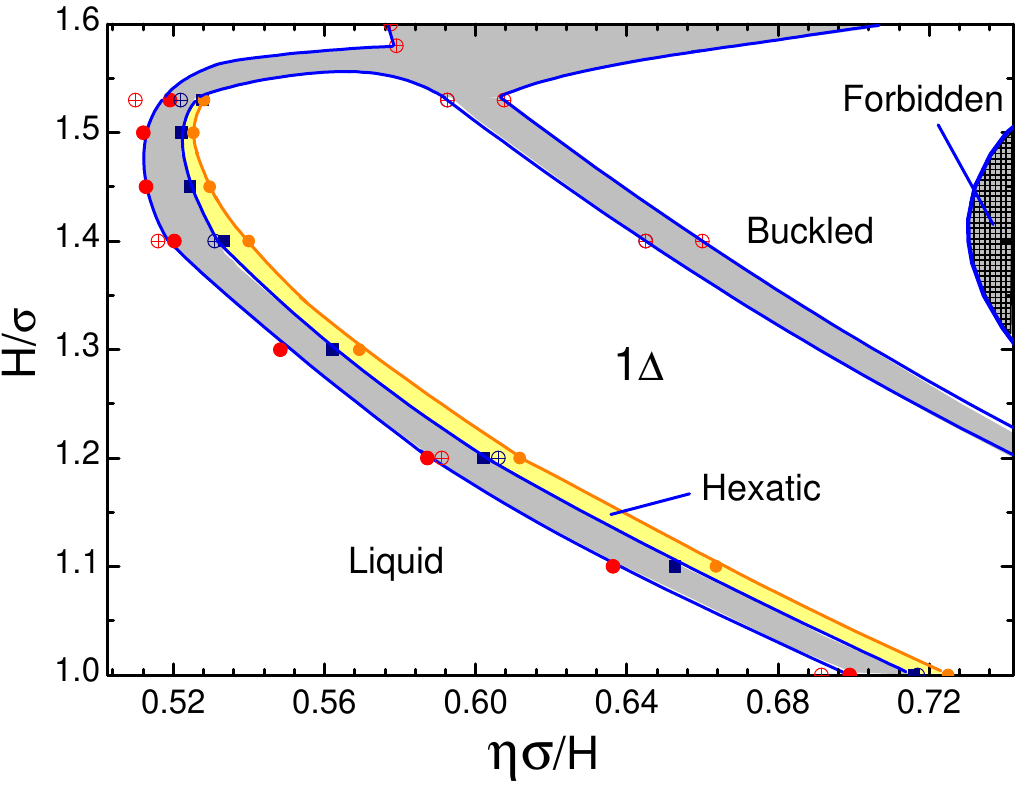}
\label{Fig:Scalinghd}
\caption{Phase diagram of hard spheres with diameter $\sigma$ confined between two parallel hard walls as a function of plate separation $H/\sigma$ and reduced density $\eta\sigma/H=\pi N \sigma^3/4AH$. We denote the stable one-phase regions with the labels "Liquid" and "Hexatic". The labels "1$\Delta$"  denotes the crystalline monolayer with triangular symmetry and the label "Buckled" correspond to the buckling phase consisting of rows that are displaced in height \cite{0953-8984-18-28-L02}. We also denote the region that is forbidden   as it exceeds the maximum possible packing fraction. The gray regions denote the two-phase coexistence regions, while the yellow region denotes the hexatic phase. The solid symbols are the phase boundaries determined in this work  and the open symbols are obtained from  Ref. \cite{0953-8984-18-28-L02}. The lines are guides to the eye.}
\end{figure}

\subsection{Finite size scaling of the positional order parameter}
To characterize the coexisting phase at high densities, we performed a sub-block scaling analysis to the 2D positional order parameter  in reciprocal space
\begin{equation}
\Psi_G = \left|\frac{1}{N} \sum_{i=1}^{N} \exp{(i \mathbf{G} \cdot \mathbf{r}_i)} \right|^2,
\end{equation}
where the sum runs over all particles $i$, $\mathbf{r}_i$ is the 2D projection of the position of particle $i$ and  $\mathbf{G}$ denotes the wave vector that corresponds to a diffraction peak and equals  $2\pi/a$ with $a$  the averaged lattice spacing. We remark here that the average lattice spacing might differ  from the lattice spacing of an ideal triangular lattice due to vacancies and other defects~\cite{PhysRevLett.107.155704}.
We calculated $\Psi_G$ for varying sub-block sizes $L_B/L$ with $L=\sqrt{L_xL_y}/4$ and analyzed the scaling of $\ln(\Psi_{\mathbf{G}}(L_b)/\Psi_{\mathbf{G}}(L))$ versus  $\ln(L_b/L)$.  According to the KTHNY theory, the positional order parameter is expected to decay algebraically, \emph{i.e.}, $\Psi_{\mathbf{G}}(L) \propto L^{-\alpha}$ with an exponent $0 \leq \alpha \leq 1/3$ in the solid phase, while in the liquid and hexatic phase the translational order decays exponentially \cite{bagchi}. As we find similar results for all plate separations $1 \leq H/\sigma \leq 1.53$, we only present  results for $H/\sigma=1.1$ and refer the reader to the Supplementary Information for other values of $H/\sigma$. In Fig. 1(b) we present the sub-block scaling analysis for $H/\sigma=1.1$.  We find that the positional order decays algebraically with a slope $\alpha<1/3$ for $\eta \geq 0.728$, which is higher than the coexisting densities $\eta_L=0.700$ and $\eta_H=0.718$ of the liquid and hexatic phase, respectively, as determined from the Maxwell construction, indicating a small density regime with a stable hexatic phase. We thus find evidence for a first-order fluid-hexatic phase transition as  supported by the Maxwell construction and a continuous hexatic-solid transition at $\eta_{HS} \simeq 0.728$.

\subsection{Positional and bond orientational correlation functions}
In order to corroborate our findings, we also computed the  positional correlation function in reciprocal space
\begin{equation}
   g_{\mathbf{G}}(r)=\langle \psi^*_{\mathbf{G}}(\mathbf{r}'+\mathbf{r})\psi_{\mathbf{G}}(\mathbf{r}')\rangle
\end{equation}
 with
$\psi_{\mathbf{G}}(\mathbf{r})=\exp{(i \mathbf{G} \cdot \mathbf{r})}$  and $\mathbf{G}$ as defined above,  and the bond orientational order correlation function
\begin{equation}
 g_{6}(r) = \langle \psi_{6}^*(\mathbf{r}'+\mathbf{r}) \psi_{6}(\mathbf{r}') \rangle
\end{equation}
 with  $\psi_6 (\mathbf{r}_i) = \frac{1}{N_i}\sum_{j\in N_i}\exp{(i6\theta_{ij})}$,
 where the sum runs over $N_i$ neighbors $j$  of particle $i$. We show  exemplarily the correlation functions in Fig. 1(c) and (d) for $H/\sigma=1.1$ at varying packing fractions $\eta$.
We observe that in the density regime  $\eta~ \in~ [0.718;0.728]$ the positional order $g_{\mathbf{G}}(r)$ decays exponentially, while the bond orientational order  is quasi-long ranged which supports again the presence of a stable hexatic phase and confirms the liquid-hexatic phase coexistence. For $\eta \sim 0.728$, the positional order decays algebraically with an exponent $-1/3$, which corresponds to a continuous solid-hexatic transition according to the KTHNY theory. In addition, we find that the positional correlation function $g_G(r)$  is well-fitted by a stretched exponential function $e^{-(r/\xi)^\beta}$ in the hexatic phase with  a  correlation length $\xi \sim 20 \sigma$ and $0.2 \leq \beta \leq 1$.

\begin{figure}
\centering
\includegraphics[width=0.5\textwidth]{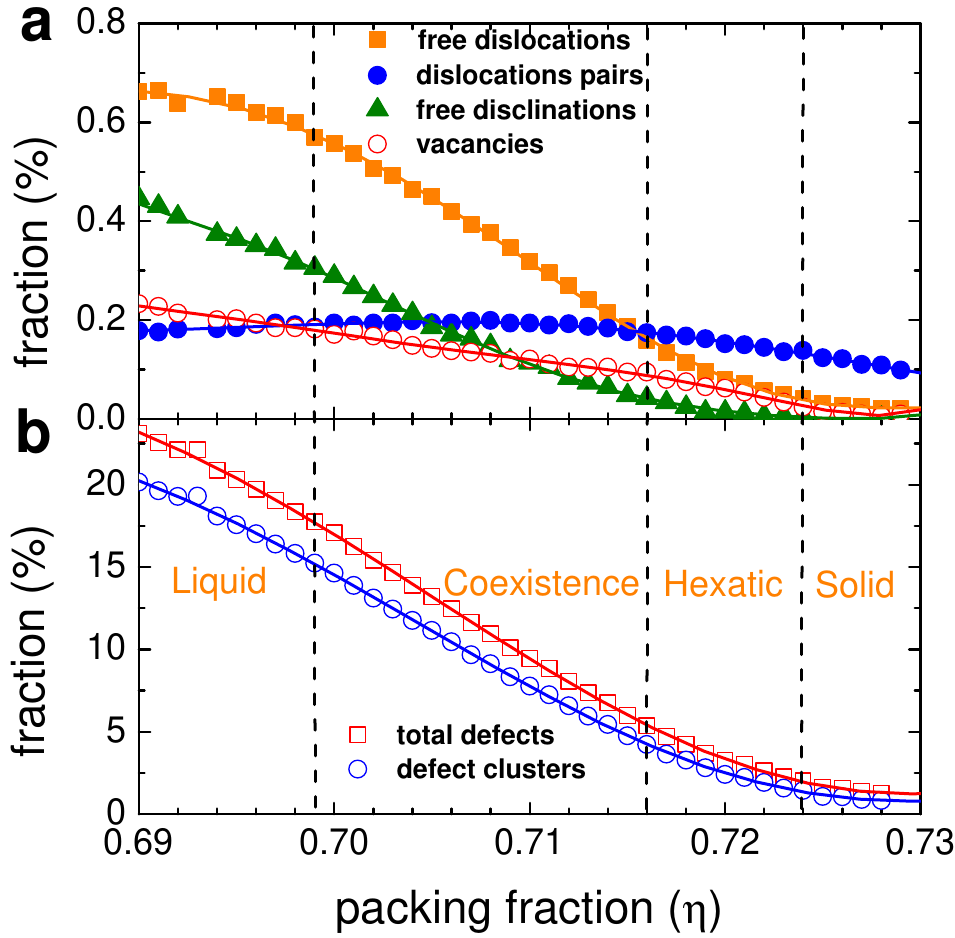}
\label{Fig:defectcoreE}
\caption{(a) Number fraction  of  free dislocations (squares), free disclinations (triangles), bound dislocation pairs (filled circles) and vacancies (open circles)  versus the 2D packing fraction $\eta=\pi N \sigma^2/4A$ for a 2D hard-disk system ($H/\sigma=1$). We define the number fraction (in $\%$) as the number of respective defects divided by the total number of particles. (b) Total number of defects per particle (open squares) and number of defects in  defect clusters per particle (open circles) versus $\eta$.}
\end{figure}

\subsection{Effect of out-of-plane fluctuations}
Employing the same analysis as described above for other values of $H/\sigma$ (see Supplementary information), we find that the two-step melting behavior persists in the whole range of plate separations $ 1 \leq H/\sigma \leq 1.53$, although for $H/\sigma=1.53$ the hexatic phase is stable only in a minute  density regime. For $H/\sigma=1.56$, we did not observe a stable crystalline monolayer with a triangular symmetry.  We present our results on the phase behavior of hard spheres confined between two parallel hard walls in Fig. 2 along with the relevant part of the phase diagram of Ref.~\cite{0953-8984-18-28-L02}. We find a remarkable agreement between the coexisting densities as obtained from the Mayer-Wood loop in the EOS and those determined using free-energy calculations for the fluid-solid transition in confined hard spheres \cite{0953-8984-18-28-L02}.  We remark  that the system sizes of 200 particles as employed in Ref. \cite{0953-8984-18-28-L02}  were too small to distinguish the hexatic phase from the solid. It is also important to note that our results on the melting transition for $H/\sigma=1$ based on a different method to distinguish the different phases, \emph{i.e.}, a sub-block scaling analysis,  matches with those obtained for a 2D hard-disk system  \cite{PhysRevLett.107.155704,arXiv1211.1645}. However, our finding of a stable hexatic phase for  plate separations as high as $H/\sigma= 1.53$ contrasts simulations of a confined  Lennard-Jones system, where a stable hexatic phase was only found when the out-of-plane particle fluctuations are less than $0.15\sigma$ \cite{gribova:054514}. We wish to remark  that the system size of $N = 212^2$ was perhaps too small
to identify the hexatic phase here. We also mention that  an integral equation theory of confined hard spheres predicts a  continuous liquid-hexatic phase transition for plate separations  $H/\sigma< 0.4$, but  a first-order liquid-solid transition for larger plate separations~\cite{PhysRevE.78.011602}. Our results thus differ from these theoretical predictions as (i) we find a first-order instead of a continuous liquid-hexatic transition for all plate separations  $ 1 \leq H/\sigma \leq 1.53$, and (ii) we do not find  a cross-over from a continuous liquid-hexatic transition to a first-order liquid-solid transition at sufficiently large plate separations. We also mention here that the authors of Ref.~~\cite{PhysRevE.78.011602} were not able to exclude a possible first-order hexatic-liquid melting transition as they employed a bifurcation analysis.

\begin{figure*}
\centering
\includegraphics[width=0.95\textwidth]{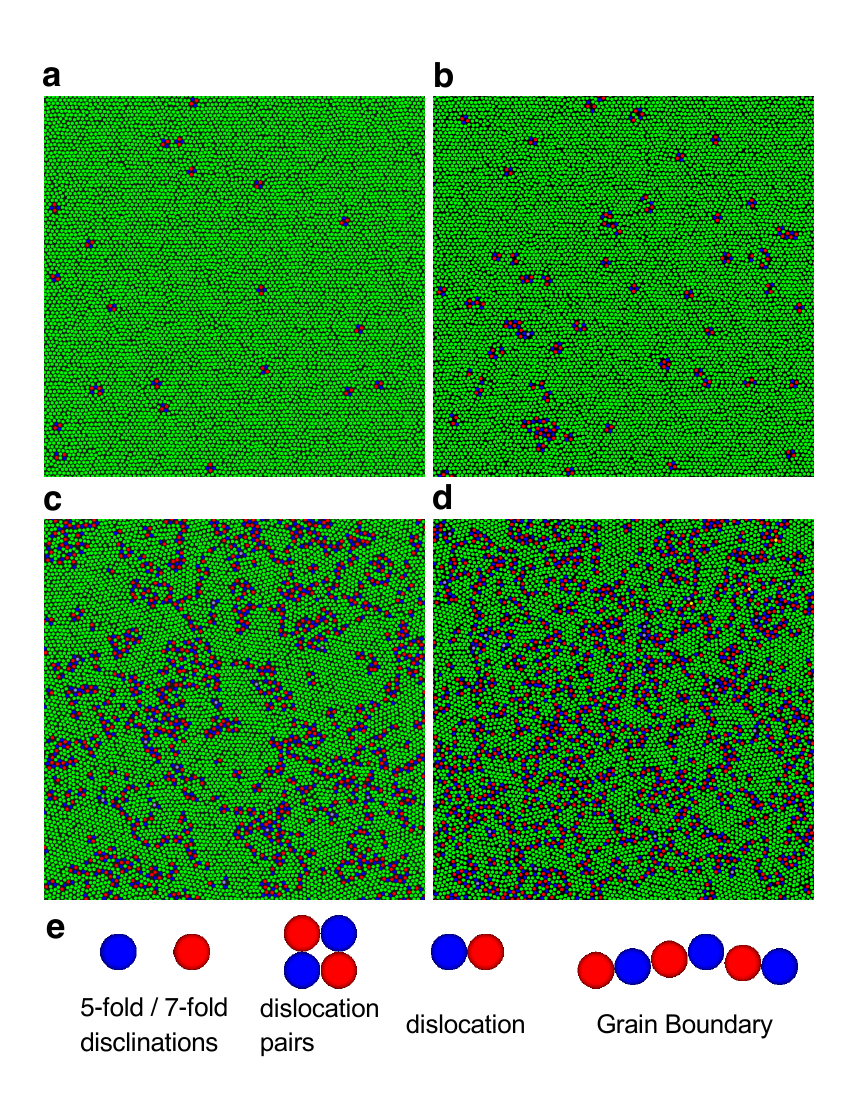}
\label{Fig:defect}
\caption{Typical configurations of hard disks ($H/\sigma=1$) in a $100 \sigma\times100\sigma$ sub-box for (a) $\eta = 0.728$ (solid), (b) $\eta = 0.722$ (hexatic), (c) $\eta = 0.718$ (coexisting hexatic phase), and (d)  $\eta = 0.699$ (coexisting liquid phase). Green particles are particles with six nearest neighbors, blue particles are $5$-fold defects, red particles are $7$-fold defects, yellow particles are $4$-fold defects and grey particles are $8$-fold defects. (e) schematic picture of $5$-fold and $7$-fold disclination, dislocation pair, isolated dislocation, and a grain boundary.}
\end{figure*}

\subsection{Topological defects and melting mechanism}
 The KTHNY theory suggests a two-step melting scenario where the unbinding of dislocation pairs ($5-7-5-7$ quartets) into free dislocations drives the  solid-hexatic transition, and the dissociation of dislocations ($5-7$ pairs) into free disclinations induces the hexatic-liquid transition. To investigate whether or not the melting is mediated by the unbinding of dislocations and disclinations, we calculated the number fractions of
topological defects as a function of $\eta$, where we define a defect as a particle that does not
have six nearest neighbors as determined by a Voronoi construction. In Fig. 3, we  present results  for only hard disks ($H/\sigma = 1$), but mention that we found similar results for other values of $ 1 \leq H/\sigma \leq 1.53$.  Fig. 3(a) and the supplementary movie show that in addition to a minute fraction of free dislocations and vacancies, the solid contains mainly  dislocation pairs  ($\simeq 0.1 \%$) that increases slightly with decreasing $\eta$.
A typical configuration of the solid phase with mainly dislocation pairs is presented in Fig. 4(a). It is also interesting to note that the dislocation pairs can move freely in the solid phase without destroying the positional and bond orientational order of the 2D lattice, see Supplementary movie. At the hexatic-solid transition, the  number fraction of free dislocations starts to increase to $\sim 0.2 \%$ with decreasing $\eta$, suggesting that the solid-hexatic transition is induced by the formation of free dislocations. However, the number fraction of dislocation pairs also remains increasing upon decreasing $\eta$ in the hexatic phase. Fig. 4 (b) shows a typical configuration of the hexatic phase. In addition, we found that the number fraction of  free dislocations, free disclinations, and vacancies increases with decreasing $\eta$ but remains always below $1 \%$ even in the liquid phase,  while the fraction of dislocation pairs seems to become constant in the coexistence region and liquid phase. More importantly, we found that many defect particles could not be identified as an isolated  dislocation pair, a free dislocation or a disclination, but were part of much larger defect clusters, which tend to be small and compact in the hexatic phase, but become string-like in the coexistence region and the liquid phase, see Fig. 3(b). In Fig. 4(c) and (d) and the supplementary movies, we present typical configuration of both the coexisting hexatic and liquid phase.  The fraction of particles that belonged to these string-like defect clusters (grain boundaries) with number fractions  as high as $20 \%$ outweighs the number fraction of bound dislocation pairs, free dislocations, and disclinations, indicating that the melting seems to be induced by a spontaneous proliferation of grain boundaries instead of unbinding of dislocations and disclinations. It is also worth mentioning that both the fluid as well as the hexatic phase exhibit clear crystalline patches consisting of a few hundred particles, which are surrounded by string-like defect clusters. In the case of the hexatic phase, the crystalline patches are still correlated and the bond orientational order is preserved, while in the liquid phase the crystalline domains seem to be uncorrelated, thereby destroying the bond orientational order. The supplementary movies show a time evolution of these defect structures.

\begin{figure}
\centering
\includegraphics[width=0.45\textwidth]{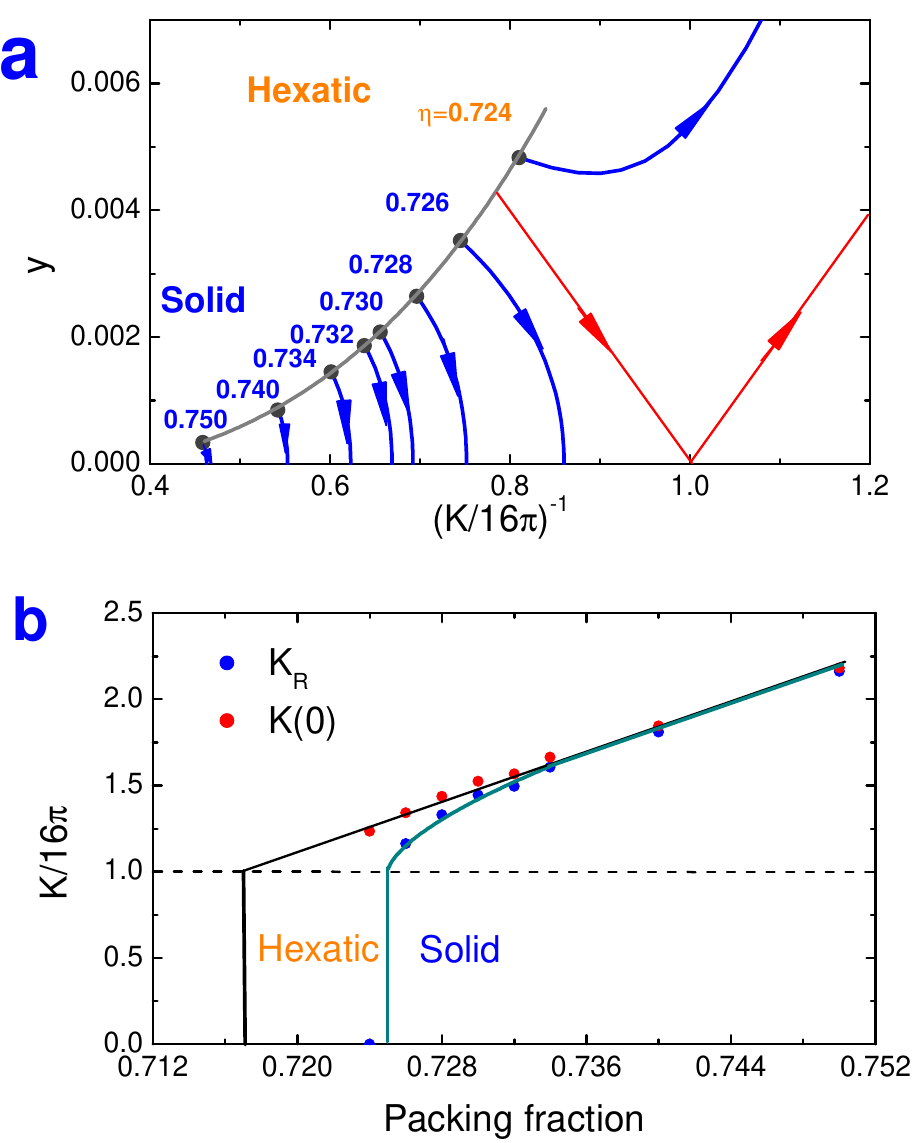}
\label{Fig:defect}
\caption{(a) Renormalization of the Young's modulus $K(l)$ and the  fugacity of dislocation pairs $y(l)$ for a system of hard disks.  Each line corresponds to a packing fraction as labeled starting from the initial values $y(0)=\exp(-E_c/k_BT)$ and the bare unrenormalized Young's modulus $K(0)$ (Gray dots). The arrows point into the direction of $l \to \infty$ corresponding to the equilibrium values of $K(l)$ and $y(l)$. The red lines indicate the flows of the renormalization group equation at the solid-hexatic transition point.  (b) Renormalized and unrenormalized Young's modulus, $K_R$ and $K$ respectively, as a function of packing fraction $eta$. Lines are guide to eye. }
\end{figure}

\subsection{Elastic constants}
According to the KTHNY theory, a continuous solid-hexatic phase transition occurs via spontaneous proliferation of dislocations when the dimensionless Young's modulus $K$ of the two-dimensional solid equals
$16 \pi$~\cite{KT73,Nelson79,Young79}:
\begin{equation}
K= \frac{8}{\sqrt{3}\rho k_BT}\frac{(\lambda+\mu)\mu}{\lambda+2\mu} = 16\pi.
\end{equation}
Here $\lambda$ and $\mu$ denote the 2D shear and bulk Lam\'e elastic constants, respectively, and $\rho=(2/\sqrt{3})a^{-2}$ is the density of the 2D triangular solid with $a$ the lattice spacing.  However, a  first-order solid-liquid transition driven by  the spontaneous proliferation of grain boundaries may preempt the solid-hexatic transition when the core energy $E_c$  of a dislocation is less than 2.84 $k_BT$ \cite{chui_prl_1982,binder_2002}. In order to compare the density at which the solid melts into an hexatic phase according to our  analysis of  the  positional and bond orientational order, we calculated the Lam\'e elastic constants from the strain fluctuations as described in Ref. ~\cite{sengupta_2000_1} and compared the results with the predictions of the KTHNY theory. To this end, we define the displacement vector ${\bf u}(t)={\bf r}(t)-{\bf R}$, where ${\bf r}(t)$ denotes the instantaneous and ${\bf R}$ the ideal lattice position of a particle. The instantaneous Lagrangian strain tensor $\epsilon_{ij}$ is then given by
\begin{equation}
\epsilon_{ij}=\frac{1}{2} \left ( \frac{\partial u_i}{\partial R_j}+\frac{\partial u_j}{\partial R_i}+\frac{\partial u_i}{\partial R_k}\frac{\partial u_k}{\partial R_j}\right ).
\end{equation}
We measure the strain fluctuations $S_{11}=S_{22}= \langle \epsilon_{xx}\epsilon_{xx} \rangle$ and $S_{12}=S_{21}= \langle \epsilon_{xx}\epsilon_{yy} \rangle$ in EDMD simulations of $N = 16384$ hard disks, and calculate the bulk and shear Lam\'e elastic coefficients using \cite{frydel_2003}
\begin{eqnarray}
\beta \mu = \frac{1}{4(S_{11}-S_{12})}-\beta P \nonumber \\
\beta \lambda = \rho \left ( \frac{\partial \beta P}{\partial \rho} \right )- \beta \mu,
\end{eqnarray}
where we used the equation of state as measured in Sec. IIIA.
The elastic constants are determined for a perfect defect-free solid without any vacancies or dislocations. A previous simulation study showed that the elastic constants are essentially unaffected by the presence of a low concentration of vacancies  \cite{PhysRevE.61.5223}. However, the elastic constant values are reduced considerably by the presence of dislocations, and should therefore be renormalized~\cite{Nelson79,Young79,sengupta_2000_1,sengupta_2000,frydel_2003}. In order to renormalize the Young's modulus by the presence of dislocations, one should first determine the core energy $E_c$ of a dislocation.
The core energy can be calculated from the probability density $p_d$ to find a dislocation pair per unit area using the relation \cite{sengupta_2000}
\begin{equation}
p_d= \frac{16\sqrt{3}\pi^2}{K-8\pi}I_0\left(\frac{K}{8\pi}\right)\exp\left(\frac{K}{8\pi}\right)\exp\left(\frac{-2E_c}{k_B  T}\right),
\end{equation}
where $I_0$ and $I_1$ are modified Bessel functions and $p_d$ is as obtained from Fig. 3a.
According to the KTHNY theory~\cite{Nelson79,Young79}, the renormalization of  the Young's modulus $K$ and the fugacity of dislocation pairs $y$ can be determined by using the recursion relations
\begin{eqnarray}
\frac{dK^{-1}(l) }{dl} &=& \frac{3}{4}\pi y^2(l) e^{\frac{K(l)}{8\pi}}\left[2I_0\left(\frac{K(l)}{8\pi}\right) - I_1\left(\frac{K(l)}{8\pi}\right)\right], \nonumber \\
\frac{dy(l)}{dl} &=& \left(2-\frac{K(l)}{8\pi}\right)y(l) + 2\pi y^2(l)e^{\frac{K(l)}{16\pi}}I_0\left(\frac{K(l)}{8\pi}\right), \nonumber
\end{eqnarray}
with $l$ the coarse-graining length scale or the renormalized flow variable.  The thermodynamic values of $K$ and $y$ are obtained in the limit $l \rightarrow \infty$.
We use the bare Young's modulus $K(l=0)$ for a defect-free solid as obtained from the strain fluctuations and $y(l=0)=e^{-E_c/k_B T}$ as the initial values for the renormalization recursion relations. In Fig. 5(a), we show the trajectories in the $y$-$K$ plane for varying packing fractions as obtained from solving the recursion relations. The arrows point into the direction of $l \rightarrow \infty$.  Fig. 5(a) shows that for $\eta > 0.724$ the dislocation fugacity $y(l) \rightarrow 0$ for $l \rightarrow \infty$, which corresponds to a solid phase without any dislocations. For $\eta = 0.724$, the dislocation fugacity $y(l) \rightarrow \infty$ for $l \rightarrow \infty$, and hence the solid melts into an hexatic phase due to a spontaneous proliferation of dislocations. In Fig. 5(b), we plot the bare $K(0)$ and renormalized Young's modulus $K_R$ as a function of $\eta$, and we find that $K_R$ changes from ~$16 \pi$ to zero at  $\eta \simeq 0.724$, thereby providing support for a KT-type solid-hexatic transition induced by the unbinding of dislocation pairs. This result agrees well with the increase in the number fraction of free dislocations at the solid-hexatic transition in Fig 4(b), and the density at which the solid-hexatic transition was predicted on the basis of the global positional order and the decay of the positional and bond orientational correlation functions as shown in Fig 1. We also note that a fit of the bare Young's modulus $K(0)$ tends to  $16\pi$ at a packing fraction $\eta \simeq 0.716$. Surprisingly, this packing fraction agrees well with the melting density of the hexatic phase as obtained from a Maxwell construction to the equation of state, where we observed a proliferation of grain boundaries. In summary, we thus find that the solid melts via unbinding of dislocation pairs into an hexatic phase via a KT-type transition, and subsequently the hexatic phase melts via a first-order transition into a fluid phase, which seems to be driven by a proliferation of grain boundaries. It is worth mentioning that in the solid phase the core energy $E_c$ of a dislocation exceeds 2.84 $k_BT$ for all packing fractions as shown in Fig. 6, and hence  a first-order fluid-solid transition does not  preempt the KT-type solid-hexatic transition. We finally note that the core energy $E_c$ can only be determined in the solid phase, where the Young's modulus remains finite (See Fig. 6). It is therefore not possible to investigate whether or not $E_c$ reaches a value of 2.84 $k_BT$ at the fluid-hexatic phase transition.

\begin{figure}
\centering
\includegraphics[width=0.45\textwidth]{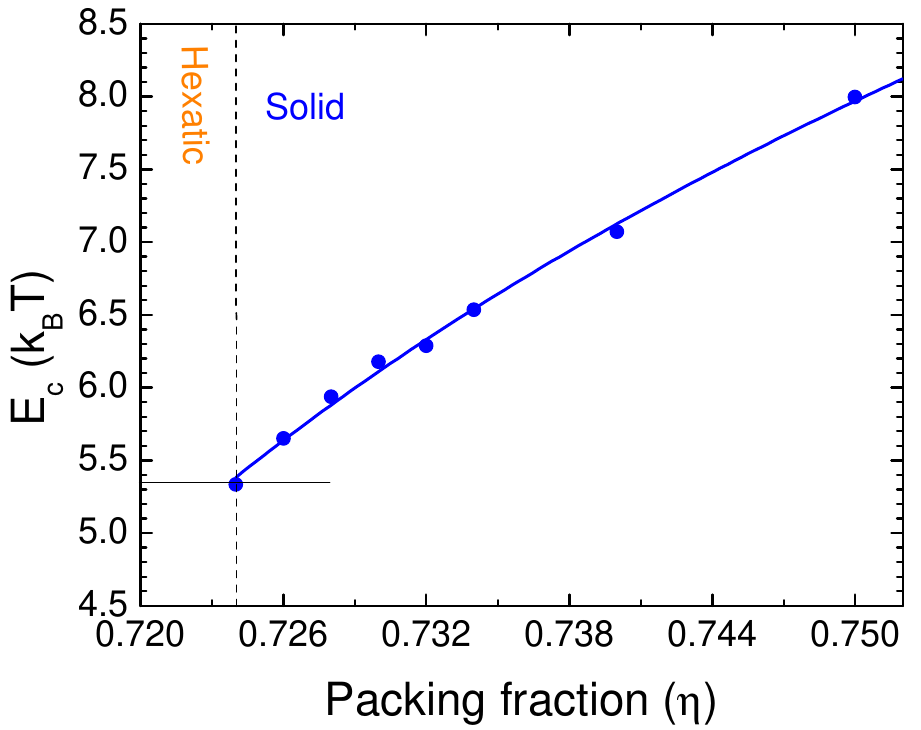}
\label{Fig:defect}
\caption{The core energy $E_c$ of a dislocation as a function of packing fraction $\eta$. Blue line denotes a fit of the data. Dashed line indicates the solid-hexatic phase transition point. Black solid line indcates the core energy $E_c > 5.35 k_BT$ for all packing fraction in solid phase. }
\end{figure}

\section{Conclusions}
In conclusion, we performed large-scale EDMD simulations of hard spheres confined between two parallel hard walls and showed that the two-stage melting scenario as observed for 2D hard disks \cite{PhysRevLett.107.155704,arXiv1211.1645} persists for quasi-2D systems of hard spheres with plate separations $ 1 \leq H/\sigma \leq 1.53$, which is of immediate importance for experiments.  Consequently, the hexatic phase sustains out-of-plane fluctuations as high as half the particle diameter, and is stable for the whole range of plate separations where a crystalline monolayer with triangular symmetry is stable. Furthermore, we show that the Young's modulus  renormalized by  dislocations, $K_R$, equals $16 \pi$ at the solid-hexatic transition in accordance with the predictions of the KTHNY theory. We thus find a KT-type solid-hexatic transition mediated by the unbinding of bound dislocation pairs. Subsequently, the KTHNY theory predicts a continous hexatic-liquid phase transition induced by the unbinding of dislocations into disclinations. However, our simulations strongly indicates that the KT-type hexatic-liquid transition is preempted by a first-order transition as demonstrated by the Mayer-Wood loops in the equation of state, and seems to be driven by the formation of string-like defect clusters or grain boundaries.

Comparing our results with experiments, we find that the available experimental data depends sensitively on the precise details of the interaction potentials between the particles. Experiments on colloidal spheres confined between two glass plates seemed to show a first-order fluid-solid transition in the case of silica spheres, but as only a few densities were studied the hexatic phase could have been missed here very easily~\cite{karnchanaphanurach_pre_2000}. For short-repulsive microgel and dipolar spheres  a liquid-hexatic and hexatic-solid transition were reported but the accuracy of the data was insufficient to determine the order of the  transitions \cite{PhysRevE.77.041406,PhysRevLett.104.205703,kusner_prl_1994}.
An alternative way to study 2D melting in colloidal systems is to adsorb  particles at air-liquid or liquid-liquid interfaces. Support has been found  for a two-stage  melting scenario for dipolar and long-range repulsive spheres~\cite{PhysRevLett.82.2721,0953-8984-1-9-015}, but the order of the transition was again not established. However, more recent experiments on dipolar spheres show compelling evidence for two consecutive continuous transitions in agreement with the KTHNY theory based on a careful analysis of the elastic constants  ~\cite{grunberg_prl_2004,keim_pre_2007,deutchlander_prl_2013}. Finally, a first-order fluid-hexatic and a continuous hexatic-solid transition was observed for particles interacting with soft repulsions~\cite{lin_jcp_2007,lin_2006}. However, the nature of the 2D melting transition for colloidal {\em hard}  spheres remains elusive, and  hence there is  an urgent need  to investigate in experiments and theory what the effect is of interparticle potentials, \emph{e.g.} range of repulsions \cite{mazars}, attractions \cite{bladon_1995,marcus_prl_1996} or  temperature-dependence, on the quasi-2D melting mechanism. However, it should be noted here that our calculations show that the hexatic phase is only stable in a minute density regime in the case of  hard spheres, which can be missed very easily in both simulations and experiments. Additionally, the order of the transition  is difficult to ascertain due to finite-size effects. We hope that our analysis, \emph{i.e.}, first establishing the coexistence region (if present) and then analyzing the decay of the positional and bond orientational order of the stable phases, can guide future experimental and simulation studies to establish the nature of the melting mechanism.

\section{Acknowledgements}
We thank R. van Roij,  J. de Graaf,  L. Filion, A. Fortini, Y. Han, and W. Krauth for helpful discussions. This work was supported by a NWO-Vici grant.


\begin{thebibliography}{52}%
\makeatletter
\providecommand \@ifxundefined [1]{%
 \@ifx{#1\undefined}
}%
\providecommand \@ifnum [1]{%
 \ifnum #1\expandafter \@firstoftwo
 \else \expandafter \@secondoftwo
 \fi
}%
\providecommand \@ifx [1]{%
 \ifx #1\expandafter \@firstoftwo
 \else \expandafter \@secondoftwo
 \fi
}%
\providecommand \natexlab [1]{#1}%
\providecommand \enquote  [1]{``#1''}%
\providecommand \bibnamefont  [1]{#1}%
\providecommand \bibfnamefont [1]{#1}%
\providecommand \citenamefont [1]{#1}%
\providecommand \href@noop [0]{\@secondoftwo}%
\providecommand \href [0]{\begingroup \@sanitize@url \@href}%
\providecommand \@href[1]{\@@startlink{#1}\@@href}%
\providecommand \@@href[1]{\endgroup#1\@@endlink}%
\providecommand \@sanitize@url [0]{\catcode `\\12\catcode `\$12\catcode
  `\&12\catcode `\#12\catcode `\^12\catcode `\_12\catcode `\%12\relax}%
\providecommand \@@startlink[1]{}%
\providecommand \@@endlink[0]{}%
\providecommand \url  [0]{\begingroup\@sanitize@url \@url }%
\providecommand \@url [1]{\endgroup\@href {#1}{\urlprefix }}%
\providecommand \urlprefix  [0]{URL }%
\providecommand \Eprint [0]{\href }%
\providecommand \doibase [0]{http://dx.doi.org/}%
\providecommand \selectlanguage [0]{\@gobble}%
\providecommand \bibinfo  [0]{\@secondoftwo}%
\providecommand \bibfield  [0]{\@secondoftwo}%
\providecommand \translation [1]{[#1]}%
\providecommand \BibitemOpen [0]{}%
\providecommand \bibitemStop [0]{}%
\providecommand \bibitemNoStop [0]{.\EOS\space}%
\providecommand \EOS [0]{\spacefactor3000\relax}%
\providecommand \BibitemShut  [1]{\csname bibitem#1\endcsname}%
\let\auto@bib@innerbib\@empty
\bibitem [{\citenamefont {Strandburg}(1988)}]{RevModPhys.60.161}%
  \BibitemOpen
  \bibfield  {author} {\bibinfo {author} {\bibfnamefont {K.~J.}\ \bibnamefont
  {Strandburg}},\ }\href {\doibase 10.1103/RevModPhys.60.161} {\bibfield
  {journal} {\bibinfo  {journal} {Rev. Mod. Phys.}\ }\textbf {\bibinfo {volume}
  {60}},\ \bibinfo {pages} {161} (\bibinfo {year} {1988})}\BibitemShut
  {NoStop}%
\bibitem [{\citenamefont {Dash}(1999)}]{RevModPhys.71.1737}%
  \BibitemOpen
  \bibfield  {author} {\bibinfo {author} {\bibfnamefont {J.~G.}\ \bibnamefont
  {Dash}},\ }\href {\doibase 10.1103/RevModPhys.71.1737} {\bibfield  {journal}
  {\bibinfo  {journal} {Rev. Mod. Phys.}\ }\textbf {\bibinfo {volume} {71}},\
  \bibinfo {pages} {1737} (\bibinfo {year} {1999})}\BibitemShut {NoStop}%
\bibitem [{\citenamefont {Gasser}(2009)}]{0953-8984-21-20-203101}%
  \BibitemOpen
  \bibfield  {author} {\bibinfo {author} {\bibfnamefont {U.}~\bibnamefont
  {Gasser}},\ }\href@noop {} {\bibfield  {journal} {\bibinfo  {journal} {J.
  Phys.: Condens. Matter}\ }\textbf {\bibinfo {volume} {21}},\ \bibinfo {pages}
  {203101} (\bibinfo {year} {2009})}\BibitemShut {NoStop}%
\bibitem [{\citenamefont {Kosterlitz}\ and\ \citenamefont
  {Thouless}(1973)}]{KT73}%
  \BibitemOpen
  \bibfield  {author} {\bibinfo {author} {\bibfnamefont {J.~M.}\ \bibnamefont
  {Kosterlitz}}\ and\ \bibinfo {author} {\bibfnamefont {D.~J.}\ \bibnamefont
  {Thouless}},\ }\href@noop {} {\bibfield  {journal} {\bibinfo  {journal} {J.
  Phys. C: Solid State Phys.}\ }\textbf {\bibinfo {volume} {6}},\ \bibinfo
  {pages} {1181} (\bibinfo {year} {1973})}\BibitemShut {NoStop}%
\bibitem [{\citenamefont {Nelson}\ and\ \citenamefont
  {Halperin}(1979)}]{Nelson79}%
  \BibitemOpen
  \bibfield  {author} {\bibinfo {author} {\bibfnamefont {D.~R.}\ \bibnamefont
  {Nelson}}\ and\ \bibinfo {author} {\bibfnamefont {B.~I.}\ \bibnamefont
  {Halperin}},\ }\href {\doibase 10.1103/PhysRevB.19.2457} {\bibfield
  {journal} {\bibinfo  {journal} {Phys. Rev. B}\ }\textbf {\bibinfo {volume}
  {19}},\ \bibinfo {pages} {2457} (\bibinfo {year} {1979})}\BibitemShut
  {NoStop}%
\bibitem [{\citenamefont {Young}(1979)}]{Young79}%
  \BibitemOpen
  \bibfield  {author} {\bibinfo {author} {\bibfnamefont {A.~P.}\ \bibnamefont
  {Young}},\ }\href {\doibase 10.1103/PhysRevB.19.1855} {\bibfield  {journal}
  {\bibinfo  {journal} {Phys. Rev. B}\ }\textbf {\bibinfo {volume} {19}},\
  \bibinfo {pages} {1855} (\bibinfo {year} {1979})}\BibitemShut {NoStop}%
\bibitem [{\citenamefont {Binder}\ and\ \citenamefont
  {S.~Sengupta}(2002)}]{binder_2002}%
  \BibitemOpen
  \bibfield  {author} {\bibinfo {author} {\bibfnamefont {K.}~\bibnamefont
  {Binder}}\ and\ \bibinfo {author} {\bibfnamefont {P.~N.}\ \bibnamefont
  {S.~Sengupta}},\ }\href@noop {} {\bibfield  {journal} {\bibinfo  {journal}
  {J. Phys.: Condens. Matter}\ }\textbf {\bibinfo {volume} {14}},\ \bibinfo
  {pages} {2323} (\bibinfo {year} {2002})}\BibitemShut {NoStop}%
\bibitem [{\citenamefont {Chui}(1982)}]{chui_prl_1982}%
  \BibitemOpen
  \bibfield  {author} {\bibinfo {author} {\bibfnamefont {S.~T.}\ \bibnamefont
  {Chui}},\ }\href {\doibase 10.1103/PhysRevLett.48.933} {\bibfield  {journal}
  {\bibinfo  {journal} {Phys. Rev. Lett.}\ }\textbf {\bibinfo {volume} {48}},\
  \bibinfo {pages} {933} (\bibinfo {year} {1982})}\BibitemShut {NoStop}%
\bibitem [{\citenamefont {Saito}(1982)}]{saito_prl_1982}%
  \BibitemOpen
  \bibfield  {author} {\bibinfo {author} {\bibfnamefont {Y.}~\bibnamefont
  {Saito}},\ }\href {\doibase 10.1103/PhysRevLett.48.1114} {\bibfield
  {journal} {\bibinfo  {journal} {Phys. Rev. Lett.}\ }\textbf {\bibinfo
  {volume} {48}},\ \bibinfo {pages} {1114} (\bibinfo {year}
  {1982})}\BibitemShut {NoStop}%
\bibitem [{\citenamefont {Zahn}\ \emph {et~al.}(1999)\citenamefont {Zahn},
  \citenamefont {Lenke},\ and\ \citenamefont {Maret}}]{PhysRevLett.82.2721}%
  \BibitemOpen
  \bibfield  {author} {\bibinfo {author} {\bibfnamefont {K.}~\bibnamefont
  {Zahn}}, \bibinfo {author} {\bibfnamefont {R.}~\bibnamefont {Lenke}}, \ and\
  \bibinfo {author} {\bibfnamefont {G.}~\bibnamefont {Maret}},\ }\href
  {\doibase 10.1103/PhysRevLett.82.2721} {\bibfield  {journal} {\bibinfo
  {journal} {Phys. Rev. Lett.}\ }\textbf {\bibinfo {volume} {82}},\ \bibinfo
  {pages} {2721} (\bibinfo {year} {1999})}\BibitemShut {NoStop}%
\bibitem [{\citenamefont {Karnchanaphanurach}\ \emph
  {et~al.}(2000)\citenamefont {Karnchanaphanurach}, \citenamefont {Lin},\ and\
  \citenamefont {Rice}}]{karnchanaphanurach_pre_2000}%
  \BibitemOpen
  \bibfield  {author} {\bibinfo {author} {\bibfnamefont {P.}~\bibnamefont
  {Karnchanaphanurach}}, \bibinfo {author} {\bibfnamefont {B.}~\bibnamefont
  {Lin}}, \ and\ \bibinfo {author} {\bibfnamefont {S.~A.}\ \bibnamefont
  {Rice}},\ }\href {\doibase 10.1103/PhysRevE.61.4036} {\bibfield  {journal}
  {\bibinfo  {journal} {Phys. Rev. E}\ }\textbf {\bibinfo {volume} {61}},\
  \bibinfo {pages} {4036} (\bibinfo {year} {2000})}\BibitemShut {NoStop}%
\bibitem [{\citenamefont {Peng}\ \emph {et~al.}(2010)\citenamefont {Peng},
  \citenamefont {Wang}, \citenamefont {Alsayed}, \citenamefont {Yodh},\ and\
  \citenamefont {Han}}]{PhysRevLett.104.205703}%
  \BibitemOpen
  \bibfield  {author} {\bibinfo {author} {\bibfnamefont {Y.}~\bibnamefont
  {Peng}}, \bibinfo {author} {\bibfnamefont {Z.}~\bibnamefont {Wang}}, \bibinfo
  {author} {\bibfnamefont {A.~M.}\ \bibnamefont {Alsayed}}, \bibinfo {author}
  {\bibfnamefont {A.~G.}\ \bibnamefont {Yodh}}, \ and\ \bibinfo {author}
  {\bibfnamefont {Y.}~\bibnamefont {Han}},\ }\href {\doibase
  10.1103/PhysRevLett.104.205703} {\bibfield  {journal} {\bibinfo  {journal}
  {Phys. Rev. Lett.}\ }\textbf {\bibinfo {volume} {104}},\ \bibinfo {pages}
  {205703} (\bibinfo {year} {2010})}\BibitemShut {NoStop}%
\bibitem [{\citenamefont {Han}\ \emph {et~al.}(2008)\citenamefont {Han},
  \citenamefont {Ha}, \citenamefont {Alsayed},\ and\ \citenamefont
  {Yodh}}]{PhysRevE.77.041406}%
  \BibitemOpen
  \bibfield  {author} {\bibinfo {author} {\bibfnamefont {Y.}~\bibnamefont
  {Han}}, \bibinfo {author} {\bibfnamefont {N.~Y.}\ \bibnamefont {Ha}},
  \bibinfo {author} {\bibfnamefont {A.~M.}\ \bibnamefont {Alsayed}}, \ and\
  \bibinfo {author} {\bibfnamefont {A.~G.}\ \bibnamefont {Yodh}},\ }\href
  {\doibase 10.1103/PhysRevE.77.041406} {\bibfield  {journal} {\bibinfo
  {journal} {Phys. Rev. E}\ }\textbf {\bibinfo {volume} {77}},\ \bibinfo
  {pages} {041406} (\bibinfo {year} {2008})}\BibitemShut {NoStop}%
\bibitem [{\citenamefont {Rice}(2009)}]{rice_CPL_2009}%
  \BibitemOpen
  \bibfield  {author} {\bibinfo {author} {\bibfnamefont {S.~A.}\ \bibnamefont
  {Rice}},\ }\href {\doibase http://dx.doi.org/10.1016/j.cplett.2009.07.059}
  {\bibfield  {journal} {\bibinfo  {journal} {Chemical Physics Letters}\
  }\textbf {\bibinfo {volume} {479}},\ \bibinfo {pages} {1 } (\bibinfo {year}
  {2009})}\BibitemShut {NoStop}%
\bibitem [{\citenamefont {Murray}\ and\ \citenamefont
  {Van~Winkle}(1987)}]{murray_prl_1987}%
  \BibitemOpen
  \bibfield  {author} {\bibinfo {author} {\bibfnamefont {C.~A.}\ \bibnamefont
  {Murray}}\ and\ \bibinfo {author} {\bibfnamefont {D.~H.}\ \bibnamefont
  {Van~Winkle}},\ }\href {\doibase 10.1103/PhysRevLett.58.1200} {\bibfield
  {journal} {\bibinfo  {journal} {Phys. Rev. Lett.}\ }\textbf {\bibinfo
  {volume} {58}},\ \bibinfo {pages} {1200} (\bibinfo {year}
  {1987})}\BibitemShut {NoStop}%
\bibitem [{\citenamefont {Kusner}\ \emph {et~al.}(1994)\citenamefont {Kusner},
  \citenamefont {Mann}, \citenamefont {Kerins},\ and\ \citenamefont
  {Dahm}}]{kusner_prl_1994}%
  \BibitemOpen
  \bibfield  {author} {\bibinfo {author} {\bibfnamefont {R.~E.}\ \bibnamefont
  {Kusner}}, \bibinfo {author} {\bibfnamefont {J.~A.}\ \bibnamefont {Mann}},
  \bibinfo {author} {\bibfnamefont {J.}~\bibnamefont {Kerins}}, \ and\ \bibinfo
  {author} {\bibfnamefont {A.~J.}\ \bibnamefont {Dahm}},\ }\href {\doibase
  10.1103/PhysRevLett.73.3113} {\bibfield  {journal} {\bibinfo  {journal}
  {Phys. Rev. Lett.}\ }\textbf {\bibinfo {volume} {73}},\ \bibinfo {pages}
  {3113} (\bibinfo {year} {1994})}\BibitemShut {NoStop}%
\bibitem [{\citenamefont {Mazars}(null)}]{mazars}%
  \BibitemOpen
  \bibfield  {author} {\bibinfo {author} {\bibfnamefont {M.}~\bibnamefont
  {Mazars}},\ }\href@noop {} {\bibfield  {journal} {\bibinfo  {journal}
  {arXiv:1301.1571}\ }\textbf {\bibinfo {volume} {null}},\ \bibinfo {pages}
  {null} (\bibinfo {year} {null})}\BibitemShut {NoStop}%
\bibitem [{\citenamefont {Marcus}\ and\ \citenamefont
  {Rice}(1996)}]{marcus_prl_1996}%
  \BibitemOpen
  \bibfield  {author} {\bibinfo {author} {\bibfnamefont {A.~H.}\ \bibnamefont
  {Marcus}}\ and\ \bibinfo {author} {\bibfnamefont {S.~A.}\ \bibnamefont
  {Rice}},\ }\href {\doibase 10.1103/PhysRevLett.77.2577} {\bibfield  {journal}
  {\bibinfo  {journal} {Phys. Rev. Lett.}\ }\textbf {\bibinfo {volume} {77}},\
  \bibinfo {pages} {2577} (\bibinfo {year} {1996})}\BibitemShut {NoStop}%
\bibitem [{\citenamefont {von Gr\"unberg}\ \emph {et~al.}(2004)\citenamefont
  {von Gr\"unberg}, \citenamefont {Keim}, \citenamefont {Zahn},\ and\
  \citenamefont {Maret}}]{grunberg_prl_2004}%
  \BibitemOpen
  \bibfield  {author} {\bibinfo {author} {\bibfnamefont {H.}~\bibnamefont {von
  Gr\"unberg}}, \bibinfo {author} {\bibfnamefont {P.}~\bibnamefont {Keim}},
  \bibinfo {author} {\bibfnamefont {K.}~\bibnamefont {Zahn}}, \ and\ \bibinfo
  {author} {\bibfnamefont {G.}~\bibnamefont {Maret}},\ }\href@noop {}
  {\bibfield  {journal} {\bibinfo  {journal} {Phys. Rev. Lett.}\ }\textbf
  {\bibinfo {volume} {93}},\ \bibinfo {pages} {255703} (\bibinfo {year}
  {2004})}\BibitemShut {NoStop}%
\bibitem [{\citenamefont {Keim}\ \emph {et~al.}(2007)\citenamefont {Keim},
  \citenamefont {Maret},\ and\ \citenamefont {von Gr\"unberg}}]{keim_pre_2007}%
  \BibitemOpen
  \bibfield  {author} {\bibinfo {author} {\bibfnamefont {P.}~\bibnamefont
  {Keim}}, \bibinfo {author} {\bibfnamefont {G.}~\bibnamefont {Maret}}, \ and\
  \bibinfo {author} {\bibfnamefont {H.}~\bibnamefont {von Gr\"unberg}},\
  }\href@noop {} {\bibfield  {journal} {\bibinfo  {journal} {Phys. Rev. E}\
  }\textbf {\bibinfo {volume} {75}},\ \bibinfo {pages} {031402} (\bibinfo
  {year} {2007})}\BibitemShut {NoStop}%
\bibitem [{\citenamefont {Deutschl\"ander}\ \emph {et~al.}(2013)\citenamefont
  {Deutschl\"ander}, \citenamefont {Horn}, \citenamefont {L\"owen},
  \citenamefont {Maret},\ and\ \citenamefont {Keim}}]{deutchlander_prl_2013}%
  \BibitemOpen
  \bibfield  {author} {\bibinfo {author} {\bibfnamefont {S.}~\bibnamefont
  {Deutschl\"ander}}, \bibinfo {author} {\bibfnamefont {T.}~\bibnamefont
  {Horn}}, \bibinfo {author} {\bibfnamefont {H.}~\bibnamefont {L\"owen}},
  \bibinfo {author} {\bibfnamefont {G.}~\bibnamefont {Maret}}, \ and\ \bibinfo
  {author} {\bibfnamefont {P.}~\bibnamefont {Keim}},\ }\href@noop {} {\bibfield
   {journal} {\bibinfo  {journal} {Phys. Rev. Lett.}\ }\textbf {\bibinfo
  {volume} {111}},\ \bibinfo {pages} {098301} (\bibinfo {year}
  {2013})}\BibitemShut {NoStop}%
\bibitem [{\citenamefont {Lin}\ and\ \citenamefont
  {Chen}(2007)}]{lin_jcp_2007}%
  \BibitemOpen
  \bibfield  {author} {\bibinfo {author} {\bibfnamefont {B.-J.}\ \bibnamefont
  {Lin}}\ and\ \bibinfo {author} {\bibfnamefont {L.-J.}\ \bibnamefont {Chen}},\
  }\href {\doibase http://dx.doi.org/10.1063/1.2409677} {\bibfield  {journal}
  {\bibinfo  {journal} {J. Chem. Phys.}\ }\textbf {\bibinfo {volume} {126}},\
  \bibinfo {pages} {034706} (\bibinfo {year} {2007})}\BibitemShut {NoStop}%
\bibitem [{\citenamefont {Jaster}(2004)}]{Jaster2004120}%
  \BibitemOpen
  \bibfield  {author} {\bibinfo {author} {\bibfnamefont {A.}~\bibnamefont
  {Jaster}},\ }\href {\doibase 10.1016/j.physleta.2004.07.055} {\bibfield
  {journal} {\bibinfo  {journal} {Physics Letters A}\ }\textbf {\bibinfo
  {volume} {330}},\ \bibinfo {pages} {120 } (\bibinfo {year}
  {2004})}\BibitemShut {NoStop}%
\bibitem [{\citenamefont {Bernard}\ and\ \citenamefont
  {Krauth}(2011)}]{PhysRevLett.107.155704}%
  \BibitemOpen
  \bibfield  {author} {\bibinfo {author} {\bibfnamefont {E.~P.}\ \bibnamefont
  {Bernard}}\ and\ \bibinfo {author} {\bibfnamefont {W.}~\bibnamefont
  {Krauth}},\ }\href {\doibase 10.1103/PhysRevLett.107.155704} {\bibfield
  {journal} {\bibinfo  {journal} {Phys. Rev. Lett.}\ }\textbf {\bibinfo
  {volume} {107}},\ \bibinfo {pages} {155704} (\bibinfo {year}
  {2011})}\BibitemShut {NoStop}%
\bibitem [{\citenamefont {Anderson}\ \emph {et~al.}(2012)\citenamefont
  {Anderson}, \citenamefont {Engel}, \citenamefont {Glotzer}, \citenamefont
  {Isobe}, \citenamefont {Bernard},\ and\ \citenamefont
  {Krauth}}]{arXiv1211.1645}%
  \BibitemOpen
  \bibfield  {author} {\bibinfo {author} {\bibfnamefont {J.~A.}\ \bibnamefont
  {Anderson}}, \bibinfo {author} {\bibfnamefont {M.}~\bibnamefont {Engel}},
  \bibinfo {author} {\bibfnamefont {S.~C.}\ \bibnamefont {Glotzer}}, \bibinfo
  {author} {\bibfnamefont {M.}~\bibnamefont {Isobe}}, \bibinfo {author}
  {\bibfnamefont {E.~P.}\ \bibnamefont {Bernard}}, \ and\ \bibinfo {author}
  {\bibfnamefont {W.}~\bibnamefont {Krauth}},\ }\href@noop {} {\bibfield
  {journal} {\bibinfo  {journal} {arXiv:1211.1645}\ }\textbf {\bibinfo {volume}
  {null}},\ \bibinfo {pages} {null} (\bibinfo {year} {2012})}\BibitemShut
  {NoStop}%
\bibitem [{\citenamefont {Alder}\ and\ \citenamefont
  {Wainwright}(1962)}]{PhysRev.127.359}%
  \BibitemOpen
  \bibfield  {author} {\bibinfo {author} {\bibfnamefont {B.~J.}\ \bibnamefont
  {Alder}}\ and\ \bibinfo {author} {\bibfnamefont {T.~E.}\ \bibnamefont
  {Wainwright}},\ }\href {\doibase 10.1103/PhysRev.127.359} {\bibfield
  {journal} {\bibinfo  {journal} {Phys. Rev.}\ }\textbf {\bibinfo {volume}
  {127}},\ \bibinfo {pages} {359} (\bibinfo {year} {1962})}\BibitemShut
  {NoStop}%
\bibitem [{\citenamefont {Hoover}\ and\ \citenamefont
  {Ree}(1968)}]{hoover_jcp_1968}%
  \BibitemOpen
  \bibfield  {author} {\bibinfo {author} {\bibfnamefont {W.~G.}\ \bibnamefont
  {Hoover}}\ and\ \bibinfo {author} {\bibfnamefont {F.~H.}\ \bibnamefont
  {Ree}},\ }\href {\doibase http://dx.doi.org/10.1063/1.1670641} {\bibfield
  {journal} {\bibinfo  {journal} {J. Chem. Phys.}\ }\textbf {\bibinfo {volume}
  {49}},\ \bibinfo {pages} {3609} (\bibinfo {year} {1968})}\BibitemShut
  {NoStop}%
\bibitem [{\citenamefont {Lee}\ and\ \citenamefont
  {Strandburg}(1992)}]{lee_prb_1992}%
  \BibitemOpen
  \bibfield  {author} {\bibinfo {author} {\bibfnamefont {J.}~\bibnamefont
  {Lee}}\ and\ \bibinfo {author} {\bibfnamefont {K.~J.}\ \bibnamefont
  {Strandburg}},\ }\href {\doibase 10.1103/PhysRevB.46.11190} {\bibfield
  {journal} {\bibinfo  {journal} {Phys. Rev. B}\ }\textbf {\bibinfo {volume}
  {46}},\ \bibinfo {pages} {11190} (\bibinfo {year} {1992})}\BibitemShut
  {NoStop}%
\bibitem [{\citenamefont {Zollweg}\ and\ \citenamefont
  {Chester}(1992)}]{PhysRevB.46.11186}%
  \BibitemOpen
  \bibfield  {author} {\bibinfo {author} {\bibfnamefont {J.~A.}\ \bibnamefont
  {Zollweg}}\ and\ \bibinfo {author} {\bibfnamefont {G.~V.}\ \bibnamefont
  {Chester}},\ }\href {\doibase 10.1103/PhysRevB.46.11186} {\bibfield
  {journal} {\bibinfo  {journal} {Phys. Rev. B}\ }\textbf {\bibinfo {volume}
  {46}},\ \bibinfo {pages} {11186} (\bibinfo {year} {1992})}\BibitemShut
  {NoStop}%
\bibitem [{\citenamefont {Weber}\ \emph {et~al.}(1995)\citenamefont {Weber},
  \citenamefont {Marx},\ and\ \citenamefont {Binder}}]{weber_prb_1995}%
  \BibitemOpen
  \bibfield  {author} {\bibinfo {author} {\bibfnamefont {H.}~\bibnamefont
  {Weber}}, \bibinfo {author} {\bibfnamefont {D.}~\bibnamefont {Marx}}, \ and\
  \bibinfo {author} {\bibfnamefont {K.}~\bibnamefont {Binder}},\ }\href
  {\doibase 10.1103/PhysRevB.51.14636} {\bibfield  {journal} {\bibinfo
  {journal} {Phys. Rev. B}\ }\textbf {\bibinfo {volume} {51}},\ \bibinfo
  {pages} {14636} (\bibinfo {year} {1995})}\BibitemShut {NoStop}%
\bibitem [{\citenamefont {Alonso}\ and\ \citenamefont
  {Fern\'andez}(1999)}]{PhysRevE.59.2659}%
  \BibitemOpen
  \bibfield  {author} {\bibinfo {author} {\bibfnamefont {J.~J.}\ \bibnamefont
  {Alonso}}\ and\ \bibinfo {author} {\bibfnamefont {J.~F.}\ \bibnamefont
  {Fern\'andez}},\ }\href {\doibase 10.1103/PhysRevE.59.2659} {\bibfield
  {journal} {\bibinfo  {journal} {Phys. Rev. E}\ }\textbf {\bibinfo {volume}
  {59}},\ \bibinfo {pages} {2659} (\bibinfo {year} {1999})}\BibitemShut
  {NoStop}%
\bibitem [{\citenamefont {Weber}\ and\ \citenamefont
  {Marx}(1994)}]{0295-5075-27-8-007}%
  \BibitemOpen
  \bibfield  {author} {\bibinfo {author} {\bibfnamefont {H.}~\bibnamefont
  {Weber}}\ and\ \bibinfo {author} {\bibfnamefont {D.}~\bibnamefont {Marx}},\
  }\href@noop {} {\bibfield  {journal} {\bibinfo  {journal} {Europhys. Lett.}\
  }\textbf {\bibinfo {volume} {27}},\ \bibinfo {pages} {593} (\bibinfo {year}
  {1994})}\BibitemShut {NoStop}%
\bibitem [{\citenamefont {Fern\'andez}\ \emph {et~al.}(1995)\citenamefont
  {Fern\'andez}, \citenamefont {Alonso},\ and\ \citenamefont
  {Stankiewicz}}]{PhysRevLett.75.3477}%
  \BibitemOpen
  \bibfield  {author} {\bibinfo {author} {\bibfnamefont {J.~F.}\ \bibnamefont
  {Fern\'andez}}, \bibinfo {author} {\bibfnamefont {J.~J.}\ \bibnamefont
  {Alonso}}, \ and\ \bibinfo {author} {\bibfnamefont {J.}~\bibnamefont
  {Stankiewicz}},\ }\href {\doibase 10.1103/PhysRevLett.75.3477} {\bibfield
  {journal} {\bibinfo  {journal} {Phys. Rev. Lett.}\ }\textbf {\bibinfo
  {volume} {75}},\ \bibinfo {pages} {3477} (\bibinfo {year}
  {1995})}\BibitemShut {NoStop}%
\bibitem [{\citenamefont {Mitus}\ \emph {et~al.}(1997)\citenamefont {Mitus},
  \citenamefont {Weber},\ and\ \citenamefont {Marx}}]{PhysRevE.55.6855}%
  \BibitemOpen
  \bibfield  {author} {\bibinfo {author} {\bibfnamefont {A.~C.}\ \bibnamefont
  {Mitus}}, \bibinfo {author} {\bibfnamefont {H.}~\bibnamefont {Weber}}, \ and\
  \bibinfo {author} {\bibfnamefont {D.}~\bibnamefont {Marx}},\ }\href {\doibase
  10.1103/PhysRevE.55.6855} {\bibfield  {journal} {\bibinfo  {journal} {Phys.
  Rev. E}\ }\textbf {\bibinfo {volume} {55}},\ \bibinfo {pages} {6855}
  (\bibinfo {year} {1997})}\BibitemShut {NoStop}%
\bibitem [{\citenamefont {Fern\'andez}\ \emph {et~al.}(1997)\citenamefont
  {Fern\'andez}, \citenamefont {Alonso},\ and\ \citenamefont
  {Stankiewicz}}]{fernandez_pre_1997}%
  \BibitemOpen
  \bibfield  {author} {\bibinfo {author} {\bibfnamefont {J.~F.}\ \bibnamefont
  {Fern\'andez}}, \bibinfo {author} {\bibfnamefont {J.~J.}\ \bibnamefont
  {Alonso}}, \ and\ \bibinfo {author} {\bibfnamefont {J.}~\bibnamefont
  {Stankiewicz}},\ }\href {\doibase 10.1103/PhysRevE.55.750} {\bibfield
  {journal} {\bibinfo  {journal} {Phys. Rev. E}\ }\textbf {\bibinfo {volume}
  {55}},\ \bibinfo {pages} {750} (\bibinfo {year} {1997})}\BibitemShut
  {NoStop}%
\bibitem [{\citenamefont {Jaster}(1999)}]{Jaster_pre_1999}%
  \BibitemOpen
  \bibfield  {author} {\bibinfo {author} {\bibfnamefont {A.}~\bibnamefont
  {Jaster}},\ }\href@noop {} {\bibfield  {journal} {\bibinfo  {journal} {Phys.
  Rev. E}\ }\textbf {\bibinfo {volume} {59}},\ \bibinfo {pages} {2594}
  (\bibinfo {year} {1999})}\BibitemShut {NoStop}%
\bibitem [{\citenamefont {Mak}(2006)}]{PhysRevE.73.065104}%
  \BibitemOpen
  \bibfield  {author} {\bibinfo {author} {\bibfnamefont {C.~H.}\ \bibnamefont
  {Mak}},\ }\href {\doibase 10.1103/PhysRevE.73.065104} {\bibfield  {journal}
  {\bibinfo  {journal} {Phys. Rev. E}\ }\textbf {\bibinfo {volume} {73}},\
  \bibinfo {pages} {065104} (\bibinfo {year} {2006})}\BibitemShut {NoStop}%
\bibitem [{\citenamefont {Gribova}\ \emph {et~al.}(2011)\citenamefont
  {Gribova}, \citenamefont {Arnold}, \citenamefont {Schilling},\ and\
  \citenamefont {Holm}}]{gribova:054514}%
  \BibitemOpen
  \bibfield  {author} {\bibinfo {author} {\bibfnamefont {N.}~\bibnamefont
  {Gribova}}, \bibinfo {author} {\bibfnamefont {A.}~\bibnamefont {Arnold}},
  \bibinfo {author} {\bibfnamefont {T.}~\bibnamefont {Schilling}}, \ and\
  \bibinfo {author} {\bibfnamefont {C.}~\bibnamefont {Holm}},\ }\href {\doibase
  10.1063/1.3623783} {\bibfield  {journal} {\bibinfo  {journal} {J. Chem.
  Phys.}\ }\textbf {\bibinfo {volume} {135}},\ \bibinfo {pages} {054514}
  (\bibinfo {year} {2011})}\BibitemShut {NoStop}%
\bibitem [{\citenamefont {Frydel}\ and\ \citenamefont
  {Rice}(2003)}]{frydel_2003}%
  \BibitemOpen
  \bibfield  {author} {\bibinfo {author} {\bibfnamefont {D.}~\bibnamefont
  {Frydel}}\ and\ \bibinfo {author} {\bibfnamefont {S.}~\bibnamefont {Rice}},\
  }\href@noop {} {\bibfield  {journal} {\bibinfo  {journal} {Phys. Rev. E}\
  }\textbf {\bibinfo {volume} {68}},\ \bibinfo {pages} {061405} (\bibinfo
  {year} {2003})}\BibitemShut {NoStop}%
\bibitem [{\citenamefont {Rapaport}(2004)}]{b3187662}%
  \BibitemOpen
  \bibfield  {author} {\bibinfo {author} {\bibfnamefont {D.~C.}\ \bibnamefont
  {Rapaport}},\ }\href@noop {} {\emph {\bibinfo {title} {The Art of Molecular
  Dynamics Simulation}}}\ (\bibinfo  {publisher} {Cambridge University Press},\
  \bibinfo {year} {2004})\BibitemShut {NoStop}%
\bibitem [{\citenamefont {Schmidt}\ and\ \citenamefont
  {L\"owen}(1996)}]{PhysRevLett.76.4552}%
  \BibitemOpen
  \bibfield  {author} {\bibinfo {author} {\bibfnamefont {M.}~\bibnamefont
  {Schmidt}}\ and\ \bibinfo {author} {\bibfnamefont {H.}~\bibnamefont
  {L\"owen}},\ }\href {\doibase 10.1103/PhysRevLett.76.4552} {\bibfield
  {journal} {\bibinfo  {journal} {Phys. Rev. Lett.}\ }\textbf {\bibinfo
  {volume} {76}},\ \bibinfo {pages} {4552} (\bibinfo {year}
  {1996})}\BibitemShut {NoStop}%
\bibitem [{\citenamefont {Fortini}\ and\ \citenamefont
  {Dijkstra}(2006)}]{0953-8984-18-28-L02}%
  \BibitemOpen
  \bibfield  {author} {\bibinfo {author} {\bibfnamefont {A.}~\bibnamefont
  {Fortini}}\ and\ \bibinfo {author} {\bibfnamefont {M.}~\bibnamefont
  {Dijkstra}},\ }\href@noop {} {\bibfield  {journal} {\bibinfo  {journal} {J.
  Phys.: Condens. Matter}\ }\textbf {\bibinfo {volume} {18}},\ \bibinfo {pages}
  {L371} (\bibinfo {year} {2006})}\BibitemShut {NoStop}%
\bibitem [{\citenamefont {Franosch}\ \emph {et~al.}(2012)\citenamefont
  {Franosch}, \citenamefont {Lang},\ and\ \citenamefont
  {Schilling}}]{franosch_2012}%
  \BibitemOpen
  \bibfield  {author} {\bibinfo {author} {\bibfnamefont {T.}~\bibnamefont
  {Franosch}}, \bibinfo {author} {\bibfnamefont {S.}~\bibnamefont {Lang}}, \
  and\ \bibinfo {author} {\bibfnamefont {R.}~\bibnamefont {Schilling}},\ }\href
  {\doibase 10.1103/PhysRevLett.109.240601} {\bibfield  {journal} {\bibinfo
  {journal} {Phys. Rev. Lett.}\ }\textbf {\bibinfo {volume} {109}},\ \bibinfo
  {pages} {240601} (\bibinfo {year} {2012})}\BibitemShut {NoStop}%
\bibitem [{\citenamefont {Mayer}\ and\ \citenamefont
  {Wood}(1965)}]{mayer_jcp_1965}%
  \BibitemOpen
  \bibfield  {author} {\bibinfo {author} {\bibfnamefont {J.~E.}\ \bibnamefont
  {Mayer}}\ and\ \bibinfo {author} {\bibfnamefont {W.~W.}\ \bibnamefont
  {Wood}},\ }\href {\doibase http://dx.doi.org/10.1063/1.1695931} {\bibfield
  {journal} {\bibinfo  {journal} {J. Chem. Phys.}\ }\textbf {\bibinfo {volume}
  {42}},\ \bibinfo {pages} {4268} (\bibinfo {year} {1965})}\BibitemShut
  {NoStop}%
\bibitem [{\citenamefont {Bagchi}\ \emph {et~al.}(1996)\citenamefont {Bagchi},
  \citenamefont {Andersen},\ and\ \citenamefont {Swope}}]{bagchi}%
  \BibitemOpen
  \bibfield  {author} {\bibinfo {author} {\bibfnamefont {K.}~\bibnamefont
  {Bagchi}}, \bibinfo {author} {\bibfnamefont {H.~C.}\ \bibnamefont
  {Andersen}}, \ and\ \bibinfo {author} {\bibfnamefont {W.}~\bibnamefont
  {Swope}},\ }\href {\doibase 10.1103/PhysRevLett.76.255} {\bibfield  {journal}
  {\bibinfo  {journal} {Phys. Rev. Lett.}\ }\textbf {\bibinfo {volume} {76}},\
  \bibinfo {pages} {255} (\bibinfo {year} {1996})}\BibitemShut {NoStop}%
\bibitem [{\citenamefont {Xu}\ and\ \citenamefont
  {Rice}(2008)}]{PhysRevE.78.011602}%
  \BibitemOpen
  \bibfield  {author} {\bibinfo {author} {\bibfnamefont {X.}~\bibnamefont
  {Xu}}\ and\ \bibinfo {author} {\bibfnamefont {S.~A.}\ \bibnamefont {Rice}},\
  }\href {\doibase 10.1103/PhysRevE.78.011602} {\bibfield  {journal} {\bibinfo
  {journal} {Phys. Rev. E}\ }\textbf {\bibinfo {volume} {78}},\ \bibinfo
  {pages} {011602} (\bibinfo {year} {2008})}\BibitemShut {NoStop}%
\bibitem [{\citenamefont {Sengupta}\ \emph
  {et~al.}(2000{\natexlab{a}})\citenamefont {Sengupta}, \citenamefont
  {Nielaba}, \citenamefont {Rao},\ and\ \citenamefont
  {Binder}}]{sengupta_2000_1}%
  \BibitemOpen
  \bibfield  {author} {\bibinfo {author} {\bibfnamefont {S.}~\bibnamefont
  {Sengupta}}, \bibinfo {author} {\bibfnamefont {P.}~\bibnamefont {Nielaba}},
  \bibinfo {author} {\bibfnamefont {M.}~\bibnamefont {Rao}}, \ and\ \bibinfo
  {author} {\bibfnamefont {K.}~\bibnamefont {Binder}},\ }\href@noop {}
  {\bibfield  {journal} {\bibinfo  {journal} {Phys. Rev. E}\ }\textbf {\bibinfo
  {volume} {61}},\ \bibinfo {pages} {1072} (\bibinfo {year}
  {2000}{\natexlab{a}})}\BibitemShut {NoStop}%
\bibitem [{\citenamefont {Bates}\ and\ \citenamefont
  {Frenkel}(2000)}]{PhysRevE.61.5223}%
  \BibitemOpen
  \bibfield  {author} {\bibinfo {author} {\bibfnamefont {M.~A.}\ \bibnamefont
  {Bates}}\ and\ \bibinfo {author} {\bibfnamefont {D.}~\bibnamefont
  {Frenkel}},\ }\href {\doibase 10.1103/PhysRevE.61.5223} {\bibfield  {journal}
  {\bibinfo  {journal} {Phys. Rev. E}\ }\textbf {\bibinfo {volume} {61}},\
  \bibinfo {pages} {5223} (\bibinfo {year} {2000})}\BibitemShut {NoStop}%
\bibitem [{\citenamefont {Sengupta}\ \emph
  {et~al.}(2000{\natexlab{b}})\citenamefont {Sengupta}, \citenamefont
  {Nielaba},\ and\ \citenamefont {Binder}}]{sengupta_2000}%
  \BibitemOpen
  \bibfield  {author} {\bibinfo {author} {\bibfnamefont {S.}~\bibnamefont
  {Sengupta}}, \bibinfo {author} {\bibfnamefont {P.}~\bibnamefont {Nielaba}}, \
  and\ \bibinfo {author} {\bibfnamefont {K.}~\bibnamefont {Binder}},\ }\href
  {\doibase 10.1103/PhysRevE.61.6294} {\bibfield  {journal} {\bibinfo
  {journal} {Phys. Rev. E}\ }\textbf {\bibinfo {volume} {61}},\ \bibinfo
  {pages} {6294} (\bibinfo {year} {2000}{\natexlab{b}})}\BibitemShut {NoStop}%
\bibitem [{\citenamefont {Armstrong}\ \emph {et~al.}(1989)\citenamefont
  {Armstrong}, \citenamefont {Mockler},\ and\ \citenamefont
  {O'Sullivan}}]{0953-8984-1-9-015}%
  \BibitemOpen
  \bibfield  {author} {\bibinfo {author} {\bibfnamefont {A.~J.}\ \bibnamefont
  {Armstrong}}, \bibinfo {author} {\bibfnamefont {R.~C.}\ \bibnamefont
  {Mockler}}, \ and\ \bibinfo {author} {\bibfnamefont {W.~J.}\ \bibnamefont
  {O'Sullivan}},\ }\href@noop {} {\bibfield  {journal} {\bibinfo  {journal} {J.
  Phys.: Condens. Matter}\ }\textbf {\bibinfo {volume} {1}},\ \bibinfo {pages}
  {1707} (\bibinfo {year} {1989})}\BibitemShut {NoStop}%
\bibitem [{\citenamefont {Lin}\ and\ \citenamefont {Chen}(2006)}]{lin_2006}%
  \BibitemOpen
  \bibfield  {author} {\bibinfo {author} {\bibfnamefont {B.-J.}\ \bibnamefont
  {Lin}}\ and\ \bibinfo {author} {\bibfnamefont {L.-J.}\ \bibnamefont {Chen}},\
  }\href {\doibase http://dx.doi.org/10.1016/j.colsurfa.2005.10.065} {\bibfield
   {journal} {\bibinfo  {journal} {Colloids and Surfaces A: Physicochemical and
  Engineering Aspects}\ }\textbf {\bibinfo {volume} {284 \-285}},\ \bibinfo
  {pages} {239 } (\bibinfo {year} {2006})}\BibitemShut {NoStop}%
\bibitem [{\citenamefont {Bladon}\ and\ \citenamefont
  {Frenkel}(1995)}]{bladon_1995}%
  \BibitemOpen
  \bibfield  {author} {\bibinfo {author} {\bibfnamefont {P.}~\bibnamefont
  {Bladon}}\ and\ \bibinfo {author} {\bibfnamefont {D.}~\bibnamefont
  {Frenkel}},\ }\href {\doibase 10.1103/PhysRevLett.74.2519} {\bibfield
  {journal} {\bibinfo  {journal} {Phys. Rev. Lett.}\ }\textbf {\bibinfo
  {volume} {74}},\ \bibinfo {pages} {2519} (\bibinfo {year}
  {1995})}\BibitemShut {NoStop}%
\end{thebibliography}
\end{document}